\documentstyle[12pt,aaspp4]{article}
\lefthead{BRITTON, GWINN & OJEDA}
\righthead{ANGULAR BROADENING OF NEARBY PULSARS}
\begin{document}
\title{Angular Broadening of Nearby Pulsars}
\author{M. C. Britton, C. R. Gwinn, and  M. J. Ojeda}
\affil{Physics Department, University of California, Santa Barbara, California
93106, USA.}
\authoraddr{mail communications to Matthew Britton at the above address, or
email mbritton@pelican.physics.ucsb.edu}
\begin{abstract}
We conducted a very long baseline interferometric (VLBI) observation of 5
nearby pulsars and did not resolve
the scattering disks of any of these sources.  Using our upper limits on the
angular diameters of these
scattering disks and published values of the broadening times and proper motion
velocities, we constrain
the possible distributions of scattering material.  The material responsible
for scattering these sources
is neither uniformly distributed nor concentrated at the surface of the Local
Bubble.  We argue that these
pulsars themselves influence their environments to produce this scattering
material.
\end{abstract}

\section{INTRODUCTION}
Fluctuations in the interstellar electron density scatter radio waves and cause
angular broadening of radio
sources.  The angular sizes of these broadened sources contain information
about the material responsible
for scattering.  Rao \& Ananthakrishnan\markcite{Rao} (1984) found that the
angular diameters of
low-latitude $(\big|b\big|<10^{\circ})$ radio sources tend to be larger in the
direction of the galactic
center.  Dennison et al\markcite{Den} (1984) found large variations in the
angular diameters of low-latitude
extragalactic radio sources along nearby lines of sight.  Some low-latitude
sources are scattered by HII
regions (Moran et al.\markcite{Moran} 1990, Molnar et al.\markcite{Mol} 1995)
or supernova remnants
(Spangler et al.\markcite{Spanb} 1986).  These results suggest that the
scattering material in the disk
of our galaxy contains both an extended component and localized features.

Pulsars are weak radio sources, but permit observation of the temporal
broadening of their pulses as well
as angular broadening.  Cordes, Weisberg, \& Boriakoff\markcite{Cordb} (1985)
found strong variations in
the temporal broadening of pulsars along nearby lines of sight.
Gwinn, Bartel, \& Cordes\markcite{Gwa} (1993) have compared the angular
broadening and temporal
broadening of pulsars to distinguish between different distributions of
scattering material.  They
conclude that the scattering properties of several distant pulsars are
consistent with a uniform
distribution of scattering material.  The Vela pulsar is a notable exception;
this young pulsar is
scattered by its supernova remnant (Desai et al.\markcite{Desai} 1992).

Because of the small scale height of the galactic disk, radio sources at high
galactic latitudes
$(\big|b\big|>10^{\circ})$ are scattered by the local interstellar medium
(ISM).
Readhead \& Hewish\markcite{Read} (1972) measured the angular diameters of
high-latitude extragalactic
radio sources and found that their angular sizes decreased smoothly with
galactic latitude.
They suggested that the scattering material is uniformly distributed above the
galactic plane with a
scale height of $\approx$ 500 pc.  In contrast, Hajivassiliou\markcite{Haj}
(1993) found in a similar
survey that a model in which all scattering occurs at the edge of the Local
Bubble could explain the
measured distribution of angular diameters at high latitude.  The Local Bubble
is a cavity of hot
coronal gas roughly 30-200 pc in extent that surrounds the earth (Cox \&
Reynolds\markcite{Cox} 1987).
Nearby pulsars tend to have stronger refractive scattering on shorter
timescales
(Gupta, Rickett, \& Coles\markcite{Gupb}, 1994) and more instances of multiple
imaging
(Gupta, Rickett, \& Lyne\markcite{Gupc} 1994) than more distant pulsars.  This
may indicate that the
distribution of material responsible for scattering differs in the two cases.
One possibility is that
nearby pulsars are scattered by the material at the surface of the Local
Bubble, while more distant
pulsars are scattered by a uniform medium.  In this paper we describe a
measurement to determine the
location of material responsible for scattering nearby pulsars.

\section{INTERFEROMETRIC MEASUREMENTS OF ANGULAR BROADENING}
The large intensity variations of pulsars in both time and frequency result
from scattering by electron
density fluctuations in the ISM (Scheuer\markcite{Sche} 1968).  Fluctuations in
the interstellar
electron density produce fluctuaions in the index of refraction, thereby
inducing variations in the
relative phases of light rays propagating along different paths.  The electric
field measured at an
antenna will vary as the phasor sum of these rays randomly coheres and
decoheres.  Rays that cohere at
frequency $\nu$ and time $t$ remain coherent over a range of frequencies
$\Delta\nu$ about $\nu$ and
over a range of times $t_{\rm{iss}}$ about $t$.  Here $\Delta\nu$ is the
scintillation bandwidth and
$t_{\rm{iss}}$ is the scintillation timescale.  An intensity maximum of
bandwidth $\Delta\nu$ and
duration $t_{\rm{iss}}$ is called a ``scintle''.

An interferometer measures the visibility, which is the complex product of the
electric fields at two
antennas.  Because the electric field at each antenna fluctuates, the
visibility
$V(\vec{x},\vec{x}+\vec{B};\omega,t)$ will be a function of the location of the
two antennas, the
frequency $\omega$, and the time $t$.   Here $\vec{x}$ and $\vec{x} + \vec{B}$
are the positions of
the antennas projected into the plane perpendicular to the line of sight to the
source.  Averaging
this visibility over position $\vec{x}$ gives
\begin{equation}
\langle V(\vec{x},\vec{x}+\vec{B};\omega,t) \rangle = V_{o}\exp \biggl[-{1
\over 2} \langle (\phi(\vec{x}) - \phi(\vec{x}+\vec{B}))^{2} \rangle \biggr]
\end{equation}
were $V_{o}$ is the average intensity measured at a single antenna and
$\phi(\vec{x})$ is the phase of
the electric field at position $\vec{x}$.  The equality follows from Mercier's
theorem.  We expect
visibility fluctuations to be a stationary random process, so that averaging
over $\vec{x}$ is
equivalent to averaging over many scintles in time or frequency.  The angular
brackets $\langle \; \rangle$
indicate both kinds of averaging.  We denote this averaged visibility as
$V(\vec{B})$.

The average visibility $V(\vec{B})$ is also the two-dimensional Fourier
transform of the sky brightness
distribution $I(\vec{q\thinspace})$, where $\vec{B}$ and $\vec{q}$ are
conjugate variables.  Expressions
for $V(\vec{B})$ have been derived for several different power spectra of
electron density fluctuations;
we use the results of Coles et al.\markcite{Coles} (1987).  For a power-law
spectrum of density
fluctuations with index $\alpha$ and inner scale $l_{i}$
\begin{equation}
V(\vec{B}) = V_{o}\exp \biggl[-{1 \over 2} \biggl({\pi \over \sqrt{2 \ln
2}}{\theta_{H} \thinspace \big|\vec{B}\big| \over \lambda} \biggr)^{\textstyle
\alpha_{\rm{vis}}-2}\biggr]
\end{equation}
where $\theta_{H}$ characterizes the angular diameter of the average scattering
disk and $\lambda$ = 92 cm
is our observing wavelength.  The index $\alpha_{\rm{vis}} = \alpha$ if
$\big|\vec{B}\big| > l_{i}$,
and $\alpha_{\rm{vis}} = 4$ if $\big|\vec{B}\big| < l_{i}$.  Estimates for the
inner scale range from
$10^{2}$ km (Spangler \& Gwinn\markcite{Spana} 1990, Molnar et al. 1995) to
$10^{6}$ km
(Coles et al.\markcite{Coles} 1987).  We take $\alpha_{\rm{vis}} = 4$ in this
paper, corresponding to a
Gaussian sky brightness distribution.  In this case $\theta_{H}$ is the full
width at half maximum (FWHM)
angular diameter of the average scattering disk.  Comparison of Equations 1 and
2 yields $\theta_{H}$:
\begin{equation}
\theta_{H} = \biggl({2 \ln 2 \over \pi^{2}}{\lambda^{2} \over
\big|\vec{B}\big|^{2}} D_{\phi}(\vec{B}\thinspace)\biggr)^{1 \over 2}
\end{equation}
where $D_{\phi}(\vec{B}\thinspace) = \langle (\phi(\vec{x}) -
\phi(\vec{x}+\vec{B}))^{2} \rangle$ is the phase structure function.

\section{OBSERVABLE CONSEQUENCES OF SCATTERING}
A comparison of angular broadening and temporal broadening provides information
on the distribution of
scattering material.  Expressions for the rms scattering angle $\theta$ and the
mean broadening time $t$
of rays propagating through a distribution of scattering material have been
derived by
Alcock \& Hatchett\markcite{Alc} (1978) and Blandford \&
Narayan\markcite{Bland} (1985).
\begin{equation}
\theta^{2} \thinspace = \thinspace {1 \over D^{2}}\int_{0}^{D}dz \thinspace
z^{2} \thinspace \psi(z)
\end{equation}
\begin{equation}
t \thinspace = \thinspace {1 \over 2cD}\int_{0}^{D}dz\thinspace z\thinspace
(D-z)\thinspace \psi(z)\thinspace
\end{equation}
Here $z$ is a coordinate from the source at $z\thinspace=\thinspace0$ to the
observer at
$z\thinspace=\thinspace D$, $\psi(z)$ is the scattering rate per unit length,
and $c$ is the speed of
light.  The rms angular diameter of the average scattering disk is $\theta$,
and is related to the FWHM
angular diameter $\theta_{H}$ by $\theta_{H} = \sqrt{\mathstrut 4 \ln 2}
\thinspace \theta$ if
$\alpha$ = 4.  The mean broadening time $t$ differs from the temporal
broadening $\tau$ by a factor
close to unity (Lee \& Jokipii\markcite{Leeb} 1975), which we neglect.  If we
assume a model of the
scattering material $\psi(z)$ with one or two parameters, we can use
measurements of $\theta_{H}$ and
$\tau$ to derive the values of these parameters.

In this paper we consider two special cases.  If the scattering material is
uniformly distributed along
the line of sight, $\psi(z)$ is a constant and the FWHM angular diameter of the
average scattering disk
is $\theta_{u} = \bigl(16 \ln 2 \thinspace c \tau / D\bigr)^{1/2}$.  If the
scattering material is
concentrated in a thin screen located a distance $d_{s}$ from earth,
$\psi(z)\propto\delta(z-D+d_{s})$
and the FWHM angular diameter of the average scattering disk is
$\theta_{s} = \bigl(8 \ln 2 c \tau \;(D-d_{s})/D d_{s} \bigr)^{1/2}$.  We
expect
$\theta_{H} = \theta_{u}$ for a pulsar scattered by a uniform medium.  If the
pulsar is
scattered by a thin screen, we can use $\theta_{H}$ and $\tau$ to compute the
distance $d_{s}$ to
the screen.

If the scattering material is concentrated in a thin screen, measurement of the
proper motion
velocity $\vec{V}_{\rm{pm}}$ of a pulsar provides an estimate of the velocity
$\vec{V}_{\rm{scr}}$
of the scattering screen.  Gupta et al.\markcite{Gupc} (1994) have discussed
this relationship,
which in our notation becomes
\begin{equation}
{\big|d_{s}\vec{V}_{\rm{pm}} - D\vec{V}_{\rm{scr}}\big|  \over D - d_{s}} =
{\lambda \over t_{\rm{iss}}} \sqrt{{D \, d_{s} \Delta\nu \over 2 \, \pi \, c \,
(D-d_{s})}}
\end{equation}
Here all velocities are measured relative to the sun and we have assumed that
the earth's orbital velocity
is negligable.  Gupta\markcite{Gupa} (1995) showed that the scattering
properties and proper motion
velocities of most pulsars are consistent with a screen halfway to the pulsar
if $\vec{V}_{\rm{scr}} = 0$.
By measuring the distance $d_{s}$ to the screen, we can estimate the velocity
$\vec{V}_{\rm{scr}}$ of the
scattering material.

\section{OBSERVATIONS AND DATA REDUCTION}
We attempted to resolve the scattering disks of 16 nearby pulsars with an
intercontinental VLBI array.
Our array consisted of telescopes at Jodrell Bank (76 m), Westerbork (phased
array), Noto (30 m), and four
25 meter telescopes from the US Very Long Baseline Array of the National Radio
Astronomy
Observatory\footnote{NRAO is operated by the Associated Univesities Inc., under
contract with the National
Science Foundation.}:  Hancock, Owens Valley, Brewster, and St. Croix.  We
observerd for 12 hours on 1994
May 25-26, recording left-circular polarization from 323 to 337 MHz with the
Mark IIIA VLBI recording
system.  The observations were broken into 13 minute scans, and alternated
between the pulsars and a
set of extragalactic radio sources used for calibration.  We detected only 2
calibration sources,
1345+125 and 2230+114, on intercontinental baselines.

The tapes were correlated with the Mark IIIA correlator at the Max-Planck
Institut f\"{u}r
Radioastronomie in Bonn, Germany.  We integrated the data over 2.5 second
intervals and correlated
over 24 lags, yielding 0.167 MHz frequency resolution.  This is sufficient to
resolve typical
scintillation bandwidths of nearby pulsars at 326 MHz (see Table 1).  Pulsar
gating allowed us to
correlate only during the pulse, yielding an improvement in the signal to noise
ratio by a factor of
2-5 (Gwinn et al.\markcite{Gwb} 1986).  The data from Hancock were discarded
due to poor recording quality.

We used Haystack Observatory's fringe fitting program FOURFIT to recover the
cross-correlation functions
from the Mark IIIA correlator output files.  These were Fourier transformed to
give the visibility
$V(\vec{x},\vec{x}+\vec{B};\omega,t)$.

The visibility data were degraded by the fractional bitshift effect.  This
effect arises from the
inability of a lag correlator to adjust the digitized data streams continuously
to compensate for
the variation of the geometrical delay $\tau_{g}(t)$.  The sawtooth error in
delay $\Delta\tau_{g}(t)$
causes a phase error $\phi_{\rm{fbs}}(\omega, t) = \omega\Delta\tau_{g}(t)$.
This error nearly averages
out for unmodulated sources, but we found that beating between
$\phi_{\rm{fbs}}(\omega, t)$ and the pulse
gate introduced errors in the phases of as much as 0.5 radians.  We computed
the corrections to the
visibility phases by reconstructing the sawtooth delay error from the Mark IIIA
correlator output and
matching it to the pulsar ephemeris.

For each baseline in each scan we removed an instrumental model from the
visibility phase.  This model
corrected for the phase errors introduced by the differing electronic
pathlengths of the seven 2 MHz
baseband converters, the geometrical delay, and the passband curvature within
the baseband converters.
Our phase correction for the geometrical delay contained 3 contributions:  the
multiband delay, the
fringe rate, and the multiband acceleration.  This last correction is not
normally removed in
interferometry, but was found to introduce errors of up to 0.5 radians on some
baselines.  The phase
passband curvature of the baseband converters, $\phi_{\rm{pb}}(\omega)$, was
computed for each baseline
by time averaging the visibility phases over the entire data set.  These
corrections were $\le$ 0.2
radians.  The resulting visibility phases were stable to 0.1 radians for our
brightest sources.

\section{RESULTS}
Although we detected fringes on all but one of the 16 pulsars we observed, only
the 5 listed in Table 1
were bright enough to resolve individual scintles.  All pulsars and quasars
showed strong fluctuations
of visibility phase with a bandwidth greater than 14 MHz.  Scattering by the
ionosphere and the
interplanetary medium caused these fluctuations.  Ananthakrishnan \&
Dennison\markcite{Anan} (1989)
saw this effect in an observation of 3C 286.  These phase fluctuations contain
information on the
strength and spectral index of electron density fluctuations in the solar wind
(Cronyn\markcite{Cron} 1972),
and will be analyzed in a subsequent paper.  We removed this effect by
subtracting a phase
$\phi_{\rm{bb}}(t)$ averaged in frequency over the 14 MHz bandwidth of our
experiment.  Since
$\Delta\nu \le$ 1 MHz for our sources we expect this not to affect the
visibility phase fluctuations
associated with interstellar scattering.

\begin{figure}[t]
\small
\figurenum{1}
\plotone{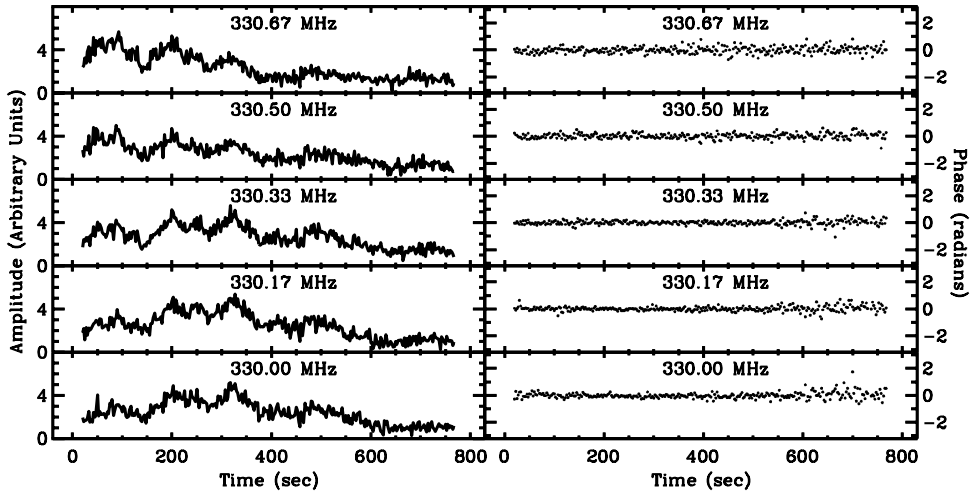}
\caption[Fig. 1]{The visibility of the pulsar B1919+21 plotted for the 7812 km
Owens Valley - Westerbork baseline.  On the left the visibility amplitude is
plotted vs. time for 5 adjacent frequency channels out of 85 observed.  The
amplitude shows variations in both time and frequency caused by scattering in
the ISM.  The visibility phase is plotted vs. time on the right, and shows no
such variations.  This indicates that the scattering disk of the pulsar is
unresolved on this baseline.  The small fluctuations in the phase are
consistent with noise.  }
\end{figure}

The visibility of the pulsar 1919+21 on the 7812 km Owens Valley - Westerbork
baseline is plotted in
Figure 1.  The visibility amplitude shows fluctuations in both time and
frequency characteristic of
scattering by the ISM.  The visibility phase shows no such variations,
indicating that the scattering
disk of the pulsar is unresolved on this baseline.  We computed the
autocorrelation function of the
real and imaginary components of the visibility, and used the single-point
offset to measure the phase
structure function (Gwinn et al.\markcite{newGwinn} (1996)).
\begin{equation}
D_{\phi} = {\big\langle
\bigl(\rm{Im}[V(\vec{x},\vec{x}+\vec{B};\omega,t)]\bigr)^{2}\big\rangle \over 4
\big\langle (\rm{Re}[V(\vec{x},\vec{x}+\vec{B};\omega,t)])^{2}\big\rangle}
\end{equation}
The single-point offset of the phase autocorrelation function suppresses noise,
but preserves the
contribution from interstellar scattering because these fluctuations correlate
over many points in
time and frequency.  We were unable to detect visibility phase fluctuations for
any of the 5 pulsars.
Table 1 lists upper limits on $D_{\phi}$ and $\theta_{H}$ and lower limits on
$d_{s}$.

In Figure 2 we plot $\theta_{H}$ vs. $\theta_{u}$ for 11 pulsars.  All
detections are taken from
Gwinn et al.\markcite{Gwa} (1993) and references therein.  The upper limits are
from Table 1 except for
the pulsar 0531+21v.  This upper limit represents the angular broadening
$\theta_{u}$ associated with the
long term variable component of the Crab pulsar's temporal broadening $\tau$.
Isaacman \& Rankin\markcite{Isa} (1977) found both a variable component and a
steady component in the
temporal broadening of this pulsar.  The variable component of $\tau$ changes
on timescales of months
to years, and is thought to be due to scattering within the Crab supernova
remnant.  The steady
component of $\tau$ arises from scattering in the ISM, and the associated
angular broadening $\theta_{u}$
is signified by 0531+21c.  The measured angular broadening of this pulsar is
consistent with scattering
in the ISM (Vandenberg\markcite{Van} 1976).

\begin{figure}[t]
\figurenum{2}
\plotone{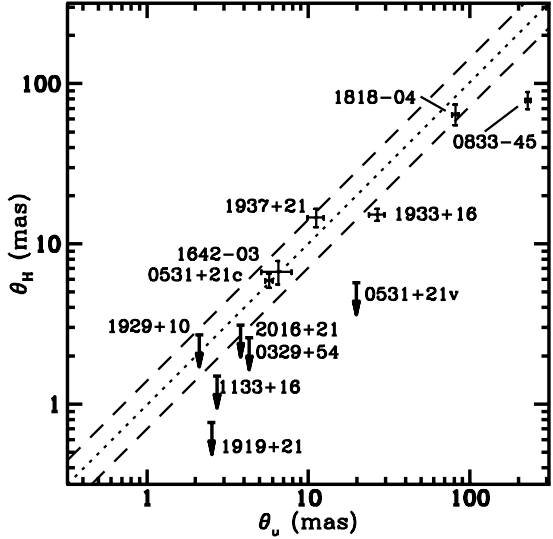}
\caption[Fig. 2]{The observed FWHM angular diameter $\theta_{H}$ of the average
scattering disk plotted vs. the  FWHM angular diameter $\theta_{u}$ expected
for scattering through a uniform medium.  If a uniform medium is responsible
for the scattering of these pulsars, points will lie on the dotted line
$\theta_{H}=\theta_{u}$.  The dashed lines delimit the range of error in
$\theta_{u}$ caused by a factor of 2 uncertainty in pulsar distances.  The
angular diameters of three nearby pulsars are inconsistent with a uniform
distribution of scattering material.}
\end{figure}

\section{DISCUSSION}
The scattering properties of most distant pulsars are consistent with
scattering through a uniform medium.
The two exceptions are the young pulsars 0531+21v and 0833-45.  These pulsars
are scattered by their
enclosing supernova remnants, and the large distance to these remnants produces
scattering disks with
smaller angular diameters.  The scattering disks of the three nearby pulsars
0329+54, 1133+16, and 1919+21
are too small to be due to scattering in a uniform medium.  If the scattering
material is concentrated in
a thin screen, the lower limits on $d_{\rm{s}}$ for the pulsars 0329+54,
1919+21, and 2016+28 preclude
scattering from the surface of the Local Bubble.  With the exception of
1133+16, all pulsars in Table 1
have galactic latitudes $\big|b\big|<5^{\circ}$.  Like low-latitude
extragalactic radio sources these
pulsars might be scattered by HII regions or supernova remnants in the galactic
disk.  However, it seems
unlikely that all of them would be scattered by small features close to the
pulsars.  We can use the
lower limit on $d_{s}$ to compute a lower limit on the component of the
velocity $\vec{V}_{\rm{scr}}$
of such a feature.  This lower limit occurs when the screen is moving in the
same direction as the pulsar.
These lower limits appear in Table 1, and indicate that the scattering material
must be moving in the
same direction as the pulsar.  While the required velocities for the scattering
material are not large,
they suggest that the scattering material may be associated with the pulsar.

To explain these results, we argue that material close to the pulsar dominates
the scattering for nearby
pulsars, producing small scattering disks.  This scattering could be caused by
strong electron density
fluctuations present at the shock front between an electron positron wind and
the ambient ISM.
Relativistic winds from pulsars can produce X-ray synchotron nebulae (Cheng \&
Helfand\markcite{Cheng} 1983,
Seward \& Wang 1988), indicating the possible presence of such a shock front.
The pulsar 2224+64 is an
example of an older high velocity pulsar that has long since left its supernova
remnant and whose wind has
evacuated a cavity in the ISM (Cordes\markcite{Corda} 1993).  These
relativistic winds may be present at
low levels in all pulsars, but may nevertheless induce the dominant electron
density fluctuations for these
nearby pulsars.

For more distant pulsars the integrated interstellar electron density
fluctuations grow to dominate the
scattering, and produce scattering disks consistent with a uniform distribution
of scattering material.
In this model the Crab pulsar would be an intermediate case, in which the ISM
dominates the angular
broadening but the effects of scattering in the supernova remnant are apparent
in the variable component
of the temporal broadening.

\acknowledgements
We thank Omer Blaes and Jon Arons for several useful discussions.  This work
was supported in part by
the National Science Foundation grant AST 92-17784.

\begin{deluxetable}{cccccccccccc}
\small
\tablecolumns{12}
\tablewidth{0pc}
\tablecaption{Scattering Properties of Nearby Pulsars\label{table1}}
\tablehead{
\colhead{Pulsar}& \colhead{$D$\tablenotemark{a}}&
\colhead{$\Delta\nu$\tablenotemark{b}}& \colhead{$\tau$\tablenotemark{c}}&
\colhead{$\theta_{u}$}& \colhead{$\big|\vec{B}\big|$}&
\colhead{$D_{\phi}^{1/2}$}& \colhead{$\theta_{H}$}&
\colhead{$d_{s}$}& \colhead{$t_{\rm{iss}}$\tablenotemark{d}}&
\colhead{$\big|\vec{V}_{\rm{pm}}\big|$\tablenotemark{d}}&
\colhead{$\big|\vec{V}_{\rm{scr}}\big|$}\\
\colhead{(B1950)}& \colhead{(pc)}& \colhead{(MHz)}& \colhead{($\mu$s)}&
\colhead{(mas)}& \colhead{(km)}&
\colhead{(rad)}& \colhead{(mas)}& \colhead{(pc)}& \colhead{(s)}&
\colhead{(km/s)}& \colhead{(km/s)}}
\startdata
0329+54& 1430&  0.028& 5.7&   4.3& 5730& $<$0.21&  $<$2.6&  $>$820&  146& 145&
$>$4\nl
1133+16&  270&   0.43& 0.37&  2.7& 7870& $<$0.17&  $<$1.5&  $>$160&   77& 475&
$>$50\nl
1919+21&  660&   0.17& 0.92&  2.5& 7810& $<$0.085& $<$0.77& $>$560&  150& 122&
$>$9\nl
1929+10&  170&   0.99& 0.16&  2.1& 7940& $<$0.30&  $<$2.7&  $>$39&   351&  86&
\nodata\nl
2016+28& 1100&  0.046& 3.5&   3.8& 6510& $<$0.28&  $<$3.1&  $>$470&  463&  12&
\nodata\nl
\enddata
\tablenotetext{a}{
Distances are from Taylor, Manchester \& Lyne\markcite{Tayb} (1993).
}
\tablenotetext{b}{
Scintillation bandwidths are from Cordes\markcite{Cord} (1986) and have been
scaled to 326 MHz using the relation $\Delta\nu \propto \nu^{4.4}.$
}
\tablenotetext{c}{
Broadening times are calculated using the relation $\tau = 1/2\pi\Delta\nu.$
}
\tablenotetext{d}{
Scintillation timescales and proper motion velocities are taken from
Gupta\markcite{Gupa} (1995).  Scintillation timescales have been scaled to 326
MHz using the relation $t_{\rm{iss}} \propto \nu^{1.2}.$
}
\end{deluxetable}

\end{document}